\documentclass[%
 aip,
 jmp,%
 amsmath,amssymb,
 reprint,%
]{revtex4-1}

\usepackage{subfigure}% Include figure files
\usepackage{graphicx}% Include figure files
\graphicspath{{./}{./images/}{./images_all/}{./images_all/spin_pumping_lambda/}{./images_all/spin_pumping_FMI/}}
\usepackage{dcolumn}% Align table columns on decimal point
\usepackage{bm}% bold math
\usepackage{xcolor}
\newcommand{\update}[1]{\textcolor{black}{#1}} 

\begin{document}

\preprint{AIP/123-QED}

\title{Determining complex spin mixing conductance and spin diffusion length from spin pumping experiments in magnetic insulator/heavy metal bilayers}% Force line breaks with \\

\author{Kuntal Roy}
\email{kuntal@iiserb.ac.in}
\noaffiliation
\affiliation{Department of Electrical Engineering and Computer Science,\\Indian Institute of Science Education and Research Bhopal, Bhopal, Madhya Pradesh 462066, India}

\begin{abstract}
Magnetic insulators are promising materials for the development of energy-efficient spintronics. Unlike metallic counterparts, the magnetic insulators are characterized by imaginary part of the interfacial spin mixing conductance as well in a bilayer with heavy metals and it is responsible for the field-like toque in spin-orbit torque devices. Here, we study the underlying theoretical constructs and develop a general strategy to determine the complex spin mixing conductance from the experimental results of ferromagnetic resonance and spin pumping. The results show that the imaginary part of the spin mixing conductance can be one order more than the real part and it matches the critical trend of spin mixing conductance with thickness of the heavy metal. The interpretation of experimental results also indicates that at small thicknesses the interface contribution becomes significant and bulk diffusion model cannot explain the results. A thickness-dependent spin diffusion length is necessary too that is tantamount to Elliott-Yafet spin relaxation mechanism in the heavy metals. Also, we effectively explain the experimental results while inserting a copper layer with varying thicknesses in between the magnetic insulator and the heavy metal using spin-circuit formalism.
\end{abstract}

%\pacs{}% PACS, the Physics and Astronomy
%\keywords{Magnetic insulators, spin pumping, complex spin mixing conductance, spin diffusion length, Elliott-Yafet spin relaxation, giant spin-orbit torque devices, spintronics}
% Use showkeys class option if keyword

\maketitle

Magnetic insulators or pure spin conductors such as yttrium iron garnet ($Y_3 Fe_5 O_{12} $, $YIG$) have attracted a lot of attention for the development of energy-efficient spintronics.~\cite{RefWorks:812,RefWorks:2593,RefWorks:1017,RefWorks:1189,RefWorks:2592,RefWorks:2596,RefWorks:2611,RefWorks:2586,RefWorks:2618,RefWorks:2625} Different phenomena such as spin-Hall magnetoresistance (SMR)~\cite{RefWorks:1024,RefWorks:1019} and spin-torque ferromagnetic resonance (ST-FMR)~\cite{RefWorks:2591} have been observed in magnetic insulators. By means of direct spin Hall effect (SHE) and a SHE layer, magnetization can be electrically exited in an adjacent magnetic insulator and the spin current can be detected by inverse spin Hall effect (ISHE) due to spin pumping.~\cite{RefWorks:992,RefWorks:885,RefWorks:2594,RefWorks:991,RefWorks:1110,RefWorks:998,RefWorks:2623} In spin pumping mechanism, a precessing magnetization emits spins into surrounding conductors and it is the reciprocal phenomenon~\cite{RefWorks:1295} of spin momentum transfer,~\cite{RefWorks:8,*RefWorks:155} according to Onsager's reciprocity.~\cite{RefWorks:1292,*RefWorks:1293} Spin pumping mechanism gives us a methodology to understand and estimate the relevant parameters in the system.~\cite{roy17_3,roy_spie_2018x} Such understandings can benefit the device design using SHE,~\cite{roy14_3,RefWorks:2588} which has potential for building future spintronic devices, alongwith other promising energy-efficient emerging devices.~\cite{roy16_spin}

The Landau-Lifshitz-Gilbert (LLG) equation~\cite{RefWorks:162,*RefWorks:161} of magnetization dynamics with phenomenological damping parameter is modified when the spin pumping contribution is considered and the experimentally observable quantities are the enhancement of damping and the ferromagnetic resonance field shift as~\cite{RefWorks:876}
\begin{equation}
\frac{d\mathbf{m}}{d\tau} = -\gamma_{eff} \mathbf{m} \times \mathbf{H_{eff}}  + \alpha_{eff} \left(\mathbf{m} \times \frac{d\mathbf{m}}
{d\tau} \right) 
\label{eq:LLG}
\end{equation}
where
\begin{equation}
\frac{\gamma}{\gamma_{eff}} = 1 - \frac{\hbar \gamma}{4 \pi M_s t_m} g_{eff,i}^{\uparrow \downarrow},
\label{eq:shift_res}
\end{equation}
\begin{equation}
\alpha_{eff} \frac{\gamma}{\gamma_{eff}} = \alpha + \alpha_{sp}, \; \alpha_{sp} = \frac{\hbar \gamma}{4 \pi M_s t_m}g_{eff,r}^{\uparrow \downarrow},
\label{eq:alpha_eff}
\end{equation}
$\mathbf{m}$ is the magnetization and $\mathbf{H_{eff}}$ is the effective field acting on the magnetization at time $\tau$, $4 \pi M_s$ is the saturation magnetization, $t_m$ is the thickness of the magnet, $\gamma_{eff}$ and $\alpha_{eff}$ are the modified (from $\gamma$ and $\alpha$, respectively) gyromagnetic ratio and damping parameter in the presence of an adjacent conductor, $\alpha_{sp}$ is the increased damping due to spin pumping, and $g_{eff}^{\uparrow \downarrow} = g_{eff,r}^{\uparrow \downarrow} + i\,g_{eff,i}^{\uparrow \downarrow}$ is the complex (reflection) \emph{effective} spin mixing conductance comprising the \emph{bare} interfacial spin mixing conductance $g^{\uparrow \downarrow} = g_r^{\uparrow \downarrow} + i\,g_i^{\uparrow \downarrow}$ and both the interface and bulk conductances of the adjacent conductor. While first principles calculations and experimental results on ferromagnetic resonance field shift show that the imaginary component of spin mixing conductance is low for metallic interfaces, magnetic insulators exhibit a significant imaginary part of the spin mixing conductance.~\cite{RefWorks:1164} The imaginary part can be conceived as an effective exchange field acting on the spin accumulation and it has immense consequence as the field-like torque in spin-orbit torque devices.~\cite{roy11_3} Magnetic proximity effect~\cite{RefWorks:1005,RefWorks:1323} has been found to be irrelevant.~\cite{RefWorks:1319,RefWorks:2605}

\begin{figure}
\centering
\includegraphics[width=0.4\textwidth]{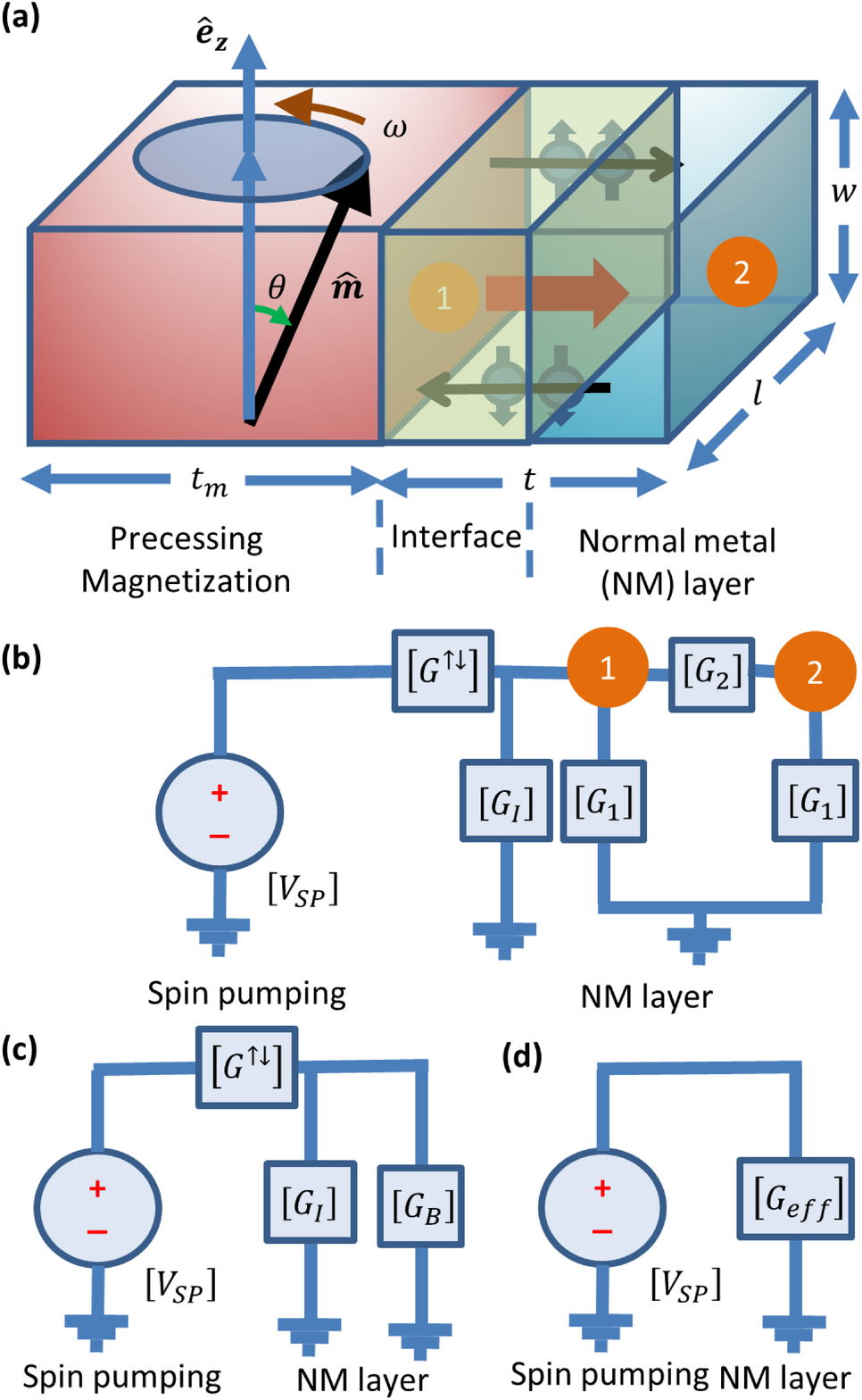}
\caption{\label{fig:spin_pumping_FMI_circuit} (a) A precessing magnetization in a magnetic insulator is pumping \emph{pure} spin current into the adjacent normal metal (NM) in a ferromagnetic resonance (FMR) experiment. Due to spin pumping, spin potentials are developed at the surfaces marked by 1 and 2. (b) In an equivalent spin-circuit diagram, the voltage source $[V_{SP}]$ acts as a spin battery, $[G^{\uparrow\downarrow}]$ is the interfacial spin mixing conductance between the magnetic layer and the NM layer, $[G_{I}]$ is the spin conductance representing the interfacial spin memory loss, and a $\pi$-network comprising the conductances $[G_1]$ and $[G_2]$ represents the bulk NM layer. (c) Minimized spin-circuit with $[G_{B}]$ representing the spin conductance due to the bulk NM layer. (d) Minimized spin-circuit with $[G_{eff}]$ representing the \emph{effective} spin mixing conductance.}
\end{figure}

\begin{figure*}
\centering
\includegraphics[width=0.9\textwidth]{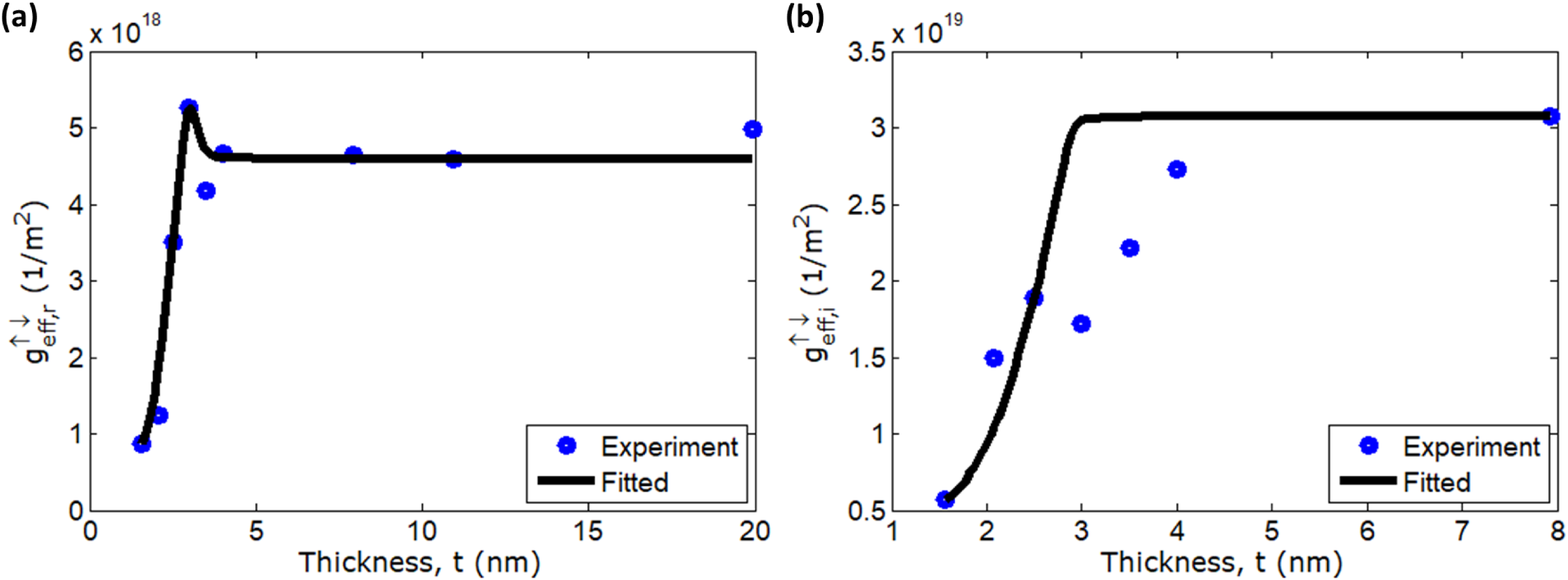}
\caption{\label{fig:spin_pumping_FMI_results} Fitting the thickness $t$ dependence of effective spin-mixing conductance, $g_{eff}^{\uparrow \downarrow}$ (a) Real component, $g_{eff,r}^{\uparrow \downarrow}$. (b) Imaginary component, $g_{eff,i}^{\uparrow \downarrow}$. Experimental data points are taken from the Ref.~\onlinecite{RefWorks:991}.}
\end{figure*}

Here, we determine the complex spin mixing conductance and spin diffusion length from the experimental results of spin pumping in magnetic insulator/heavy metal bilayers presented in Ref.~\onlinecite{RefWorks:991}. Usually, the imaginary part of the spin mixing conductance is ignored for magnetic insulators,~\cite{RefWorks:991,RefWorks:885,RefWorks:998} however, the interpretation of the experimental results show that the imaginary part of the spin mixing conductance can be one order more than the real part. Also, it turns out to be crucial to match the trend of the real part of spin mixing conductance. As the determination of the spin diffusion length ($\lambda$) in the heavy metals is concerned, recent calculations~\cite{roy17_3,roy_spie_2018x,RefWorks:2589,RefWorks:2590,RefWorks:2609} signify the Elliott-Yafet (EY) spin relaxation mechanism in which $\lambda$ is dependent on thickness ($t$) since conductivity ($\sigma$) varies with thickness of the sample.~\cite{RefWorks:885} In particular, when considering the voltage due to ISHE,~\cite{RefWorks:986,RefWorks:998} it clearly depicts the EY spin relaxation~\cite{roy17_3} and not the Dyakonov-Perel (DP)~\cite{RefWorks:2620} spin relaxation mechanism.~\cite{RefWorks:1287,RefWorks:983,RefWorks:980} The controversy acclaimed in the Ref.~\onlinecite{RefWorks:986} that two different constant values of spin diffusion lengths are required to explain the experimental results of $g_{eff}^{\uparrow \downarrow}$ and a quantity dependent on $V_{ISHE}$ can be solved by considering a thickness-dependent $\lambda$.~\cite{roy17_3} There is significant interface contribution at the magnetic insulator/heavy metal interface as well, which introduces more scattering for thin samples and lowers the conductivity. The interpretation of experimental results clearly indicate that at small thicknesses bulk diffusion model cannot explain such lowering of the conductivity. We employ spin-circuit representation of spin pumping~\cite{roy17_2} to interpret the experimental results effectively, in particular for multilayers including a sandwiched copper layer. \update{Ref.~\onlinecite{roy17_2} reproduces the standard mathematical results of spin-pumping in the literature, however, it did not consider the interface contribution.} \update{Note that magnon diffusion length is not considered since we are not dealing with propagating spin waves in magnetic insulators here.}

Figure~\ref{fig:spin_pumping_FMI_circuit}(a) shows a schematic diagram of spin pumping by a precessing magnetization in a magnetic insulator into a normal metal layer having a length $l$, width $w$, and thickness $t$. The corresponding spin-circuit representation~\cite{roy17_3} containing voltage source and conductances is shown in the Fig.~\ref{fig:spin_pumping_FMI_circuit}(b). The voltage source $[V_{SP}]$ that acts as a spin battery due to a precessing magnetization $\mathbf{m}$ and the spin current due to spin pumping can be written, respectively as
\begin{equation}
\mathbf{{V}_{SP}} = \frac{\hbar}{2e} \left(\mathbf{m} \times \frac{d\mathbf{m}}{d\tau}\right),
\label{eq:Vsp}
\end{equation}
\begin{equation}
\mathbf{{I}_{SP}} = \cfrac{\hbar}{2e}\, \left(2G_r\, \mathbf{m} \times \frac{d\mathbf{m}}{d\tau} + 2G_i\, \frac{d\mathbf{m}}{d\tau} \right).
\label{eq:Isp}
\end{equation}
The $[G^{\uparrow\downarrow}]$ is the interfacial \emph{bare} spin mixing conductance between the magnetic layer and the NM layer, which can be represented in $(\mathbf{m},d\mathbf{m}/d\tau,\mathbf{m} \times d\mathbf{m}/d\tau)$ basis as
\begin{equation}
\left\lbrack G^{\uparrow\downarrow}\right\rbrack
=\left\lbrack\begin{array}{crr}
  0 & 0 & 0\\
	0 & 2G_{r} & 2G_{i}\\ 
  0 & -2G_{i} & 2G_{r}
\end{array} \right\rbrack
\label{eq:G_SP}
\end{equation}
where $G_{r(i)} = lw\,(e^2/h)\,g_{r(i)}^{\uparrow \downarrow}$. $[G_I]$ is the spin conductance due to interfacial spin memory loss~\cite{RefWorks:1339,RefWorks:1042,RefWorks:1325,RefWorks:1111,RefWorks:1338,RefWorks:1316,RefWorks:1114,RefWorks:1319,RefWorks:1113,RefWorks:2612,RefWorks:1016,RefWorks:2603,RefWorks:2599,*RefWorks:2600,RefWorks:1351,RefWorks:2613,RefWorks:2621,RefWorks:2619,RefWorks:2598,RefWorks:2602} with parameter $\delta$ representing the spin flip probability $1-e^{-\delta}$ at the interface, where $G_I=G_{I,r}+i\,G_{I,i}$, $G_{I,r(i)} = lw\,(\delta/R_{r(i)}^*)\,sinh(\delta) = (2e^2/h)\, g_{I,r(i)}$, and $R^*=R_r^*+i\,R_i^*$ is an effective interface resistance depending on the interface spin polarization.~\cite{RefWorks:1319} Note that $R^*$ depends on the conductivity $\sigma$ and therefore thickness $t$ of the NM layer.~\cite{RefWorks:1319}

The bulk diffusion in the NM layer can be represented by a $\pi$-circuit \update{(which comes after solving the spin diffusion equation in normal metals)} as shown in the Fig.~\ref{fig:spin_pumping_FMI_circuit}(b).~\cite{roy17_2} In the absence of spin memory loss $\left\lbrack G_{1(2)} \right\rbrack=G_{1(2)} \left\lbrack I_{3\times3} \right\rbrack$, where $G_{1}=G_\lambda tanh (t/2\lambda)$, $G_{2}=G_\lambda csch (t/\lambda)$, $G_\lambda = \sigma l w/\lambda$, and $\left\lbrack I_{3\times3}\right\rbrack$ is the ${3\times3}$ identity matrix. All the bulk conductances will be altered by the spin accumulation in the NM layer with a multiplication factor $cosh(\delta)$ in the presence of spin memory loss~\cite{roy17_3} giving $\left\lbrack G_{B} \right\rbrack=G_B \left\lbrack I_{3\times3} \right\rbrack$, where $G_{B} = G_1 + G_1 G_2/(G_1 + G_2) = G_\lambda cosh(\delta) \, tanh(t/\lambda) = (2e^2/h)\, g_{B}$, as shown in Fig.~\ref{fig:spin_pumping_FMI_circuit}(c). Fig.~\ref{fig:spin_pumping_FMI_circuit}(d) shows the \emph{effective} spin mixing conductance $\left\lbrack G_{eff} \right\rbrack$ of the system where $G_{eff,r(i)} = lw\,(2e^2/h)\,g_{eff,r(i)}^{\uparrow \downarrow}$.

\begin{figure}
\centering
\includegraphics[width=0.4\textwidth]{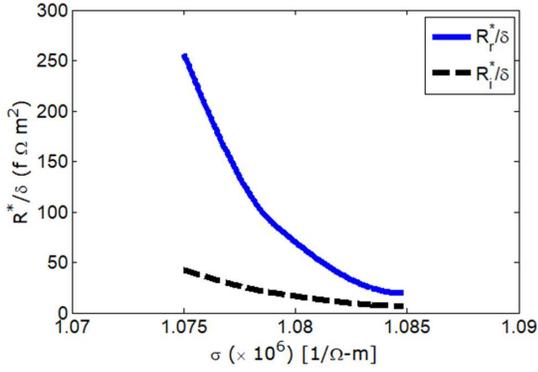}
\caption{\label{fig:spin_pumping_FMI_RI} $R_r^*/\delta$ and $R_i^*/\delta$ with the conductivity of the $Pt$ layer for $\delta=3.83$, used to fit the results in the Fig.~\ref{fig:spin_pumping_FMI_results}.}
\end{figure}

\begin{figure}
\centering
\includegraphics[width=0.4\textwidth]{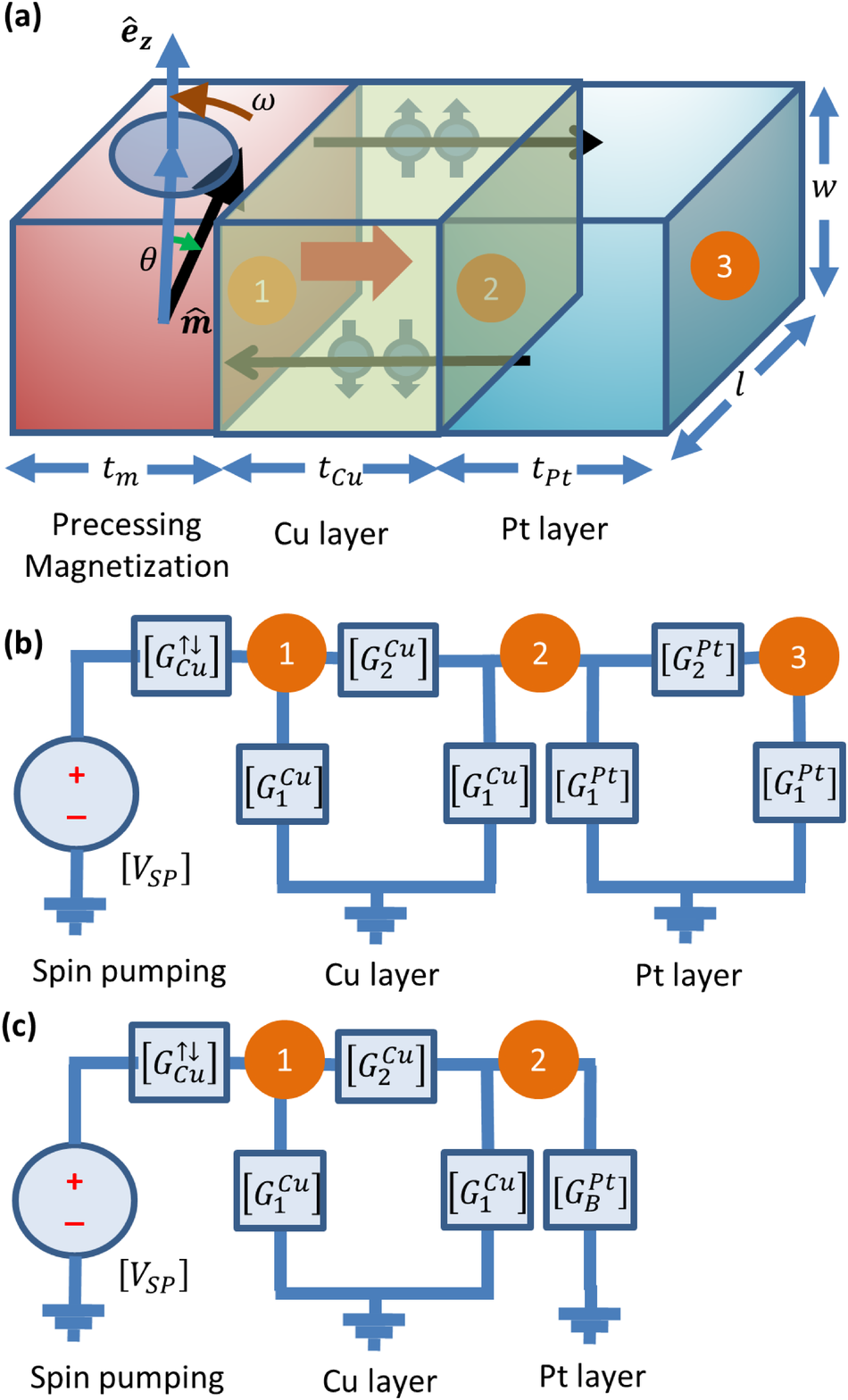}
\caption{\label{fig:spin_pumping_FMI_Cu_circuit} (a) A precessing magnetization in a magnetic insulator is pumping \emph{pure} spin current into a multilayer of Cu and Pt in a ferromagnetic resonance (FMR) experiment. There are three surfaces marked by 1,2, and 3 where spin potentials are developed. (b) In an equivalent spin circuit diagram, the voltage source $[V_{SP}]$ acts as a spin battery, $[G^{\uparrow\downarrow}_{Cu}]$ is the interfacial spin mixing conductance between the magnetic layer and the Cu layer, a $\pi$-network comprising the conductances $[G_1^{Cu}]$ and $[G_2^{Cu}]$ to represent the Cu layer, and another $\pi$-network comprising the conductances $[G_1^{Pt}]$ and $[G_2^{Pt}]$ to represent the Pt layer. (c) Minimized spin-circuit with $[G_{B}^{Pt}]$ representing the spin conductance due to the Pt layer. This spin circuit can be further minimized with the conductances calculated by series-parallel combination to get the \emph{effective} spin mixing conductance of the system.}
\end{figure}

\begin{figure}
\centering
\includegraphics[width=0.45\textwidth]{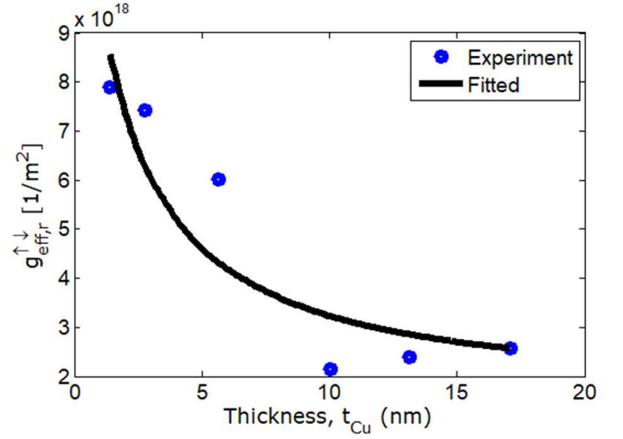}
\caption{\label{fig:spin_pumping_FMI_Cu_results} Thickness $t_{Cu}$ dependence of effective spin-mixing conductance $g_{eff,r}^{\uparrow \downarrow}$ corresponding to Fig.~\ref{fig:spin_pumping_FMI_Cu_circuit}. Experimental data points are taken from the Ref.~\onlinecite{RefWorks:991}.}
\end{figure}

From the spin-circuit in the Fig.~\ref{fig:spin_pumping_FMI_circuit}, we can write
\begin{align}
\left\lbrack\begin{array}{rr}
  G_{eff,r} & G_{eff,i} \\ 
  -G_{eff,i} & G_{eff,r}
\end{array} \right\rbrack^{-1}
&=
\left\lbrack\begin{array}{rr}
  G_r & G_i \\ 
  -G_i & G_r
\end{array} \right\rbrack^{-1} \nonumber\\
&+
\left\lbrack\begin{array}{cc}
  G_{I,r}+G_B & G_{I,i} \\ 
  -G_{I,i} & G_{I,r}+G_B
\end{array} \right\rbrack^{-1}
\label{eq:geff_mix_IB}
\end{align}
and therfore
\begin{align}
G_{eff,r(i)} = \frac{G_r^n+G_{IB,r(i)}^n}{(G_r^n+G_{IB,r}^n)^2+(G_i^n+G_{IB,i}^n)^2}
\label{eq:geff}
\end{align}
where
\begin{align}
G_{r(i)}^n = \frac{G_{r(i)}}{G_r^2+G_i^2},
\label{eq:g_norm}
\end{align}
\begin{align}
G_{IB,r}^n &= \frac{G_{I,r}+G_B}{(G_{I,r}+G_B)^2+G_{I,i}^2},
\label{eq:gIB_norm_r}
\end{align}
\begin{align}
G_{IB,i}^n &= \frac{G_{I,i}}{(G_{I,r}+G_B)^2+G_{I,i}^2}.
\label{eq:gIB_norm_i}
\end{align}
From Equation~\eqref{eq:geff_mix_IB}, we can write
\begin{align}
G_{r(i)}^n = G_{eff,r(i)}^n - G_{IB,r(i)}^n
\label{eq:g}
\end{align}
where
\begin{align}
G_{eff,r(i)}^n = \frac{G_{eff,r(i)}}{G_{eff,r}^2+G_{eff,i}^2}.
\label{eq:geff_norm}
\end{align}
From Equation~\eqref{eq:g_norm}, we can write ${G_r^n}^2 + {G_i^n}^2 = 1/(G_r^2+G_i^2)$ and thus
\begin{align}
G_{r(i)} = \frac{G_{r(i)}^n}{{G_r^n}^2+{G_i^n}^2}
\label{eq:g_norm_back}
\end{align}
which allows us to calculate the \emph{bare} spin mixing conductance $[G^{\uparrow \downarrow}]$ as given in the Equation~\eqref{eq:G_SP}.

Note that $g_{eff,r}^{\uparrow \downarrow}$ and $g_{eff,i}^{\uparrow \downarrow}$ are thickness-dependent due to the thickness dependence of $g_{I,r}$, $g_{I,i}$, and $g_{B}$. The trend of the dependence depends on how $t/\lambda$ scales with lowering thickness $t$, \update{since $\sigma$ also decreases with decreasing thickness}, with $\lambda(t) \propto \sigma(t)$, according to EY spin relaxation mechanism. It is possible to measure experimentally both the $g_{eff,i}^{\uparrow \downarrow}$  and $g_{eff,r}^{\uparrow \downarrow}$ (from Equations~\eqref{eq:shift_res} and~\eqref{eq:alpha_eff}, respectively), and conductivity $\sigma$ with thickness $t$ of the NM layer. Then choosing a value of $\lambda_{max}$ (at the highest fabricated thickness) and $(R_{r(i)}^*,\delta)$ representing interface conductance, the $g^{\uparrow \downarrow}$ can be calculated from the Equation~\eqref{eq:g_norm_back}. Using $\lambda_{max}$, $g^{\uparrow \downarrow}$, $(R_{r(i)}^*,\delta)$, and the relation $\lambda(t) \propto \sigma(t)$, we can calculate $g_{eff}^{\uparrow \downarrow}(t)$ from the Equation~\eqref{eq:geff}. We can choose the $\lambda_{max}$ and $(R_{r(i)}^*,\delta)$ that give us the best fit with the experimental data.
% and the corresponding $g^{\uparrow \downarrow}$ to characterize the experimental results of $g_{eff,r(i)}^{\uparrow \downarrow}(t)$. 

Figure~\ref{fig:spin_pumping_FMI_results} shows the fitting of the experimental results presented in the Ref.~\onlinecite{RefWorks:991}. This fitting is for $\delta=3.83$, $R_{r(i)}^*$ in Fig.~\ref{fig:spin_pumping_FMI_RI}, $\lambda_{max}=1.24\,nm$, and $(g_r^{\uparrow \downarrow},g_i^{\uparrow \downarrow})=(2.95e18,3.11e19) \,m^{-2}$. Note that $g_i^{\uparrow \downarrow}$ is one order more than $g_r^{\uparrow \downarrow}$ and this dictates the trend that there is a peak of $g_{eff,r}^{\uparrow \downarrow}$ at $t=3\,nm$. Such trend cannot be achieved for metallic magnets where there is no significant $g_i^{\uparrow \downarrow}$.~\cite{roy17_3} Fig.~\ref{fig:spin_pumping_FMI_RI} shows the trend of $R_{r(i)}^*/\delta$ with conductivity $\sigma$ when interface plays the key role ($t \leq 2.5\,nm$). The plot of $\lambda(t) \propto \sigma(t)$ is given in the Supplementary Fig.~S1. The experimental measurements of $\sigma$ with thickness $t$ is taken from Ref.~\onlinecite{RefWorks:885}. Using Kittel formula $f=\gamma \sqrt{H_r(H_r + 4\pi M_s)}$, with $f = 9.5\,GHz$, $H_r = 2545\,Oe$, and $\gamma = 2.8\,MHz/Oe$, the saturation magnetization is calculated as $4\pi M_s=1978\, G$.~\cite{RefWorks:885} Then, $\gamma_{eff}$ for different thicknesses are calculated from the FMR field shifts~\cite{RefWorks:991} and accordingly, $g_{eff,i}^{\uparrow \downarrow}$ for different thicknesses are determined from the Equation~\eqref{eq:shift_res}.

Figure~\ref{fig:spin_pumping_FMI_Cu_circuit} shows the spin-circuit while incorporating a copper (Cu) layer in between the magnetic insulator (YIG) and platinum (Pt) layers. The circuit elements are similar to as in the Fig.~\ref{fig:spin_pumping_FMI_circuit}. The thickness of Pt layer (23 nm) is quite high compared to Pt's spin diffusion length, so the $Cu|Pt$ interface contribution is ignored and due to negligible spin-orbit coupling, the interface resistance of $YIG|Cu$ is not considered. Since no significant FMR field shift is observed for this multilayer,~\cite{RefWorks:991} we neglect the imaginary spin mixing conductance and match the trend of experimental results of only $g_{eff,r}^{\uparrow \downarrow}$ in the Fig.~\ref{fig:spin_pumping_FMI_Cu_results}. The interfacial spin mixing conductance turns out to be $g_r^{\uparrow \downarrow}=3.45e19\,m^{-2}$ and the plot of $\lambda_{Cu}(t_{Cu}) \propto \sigma_{Cu}(t_{Cu})$ is given in the Supplementary Fig.~S2. The conductance $G_1^{Cu}$ is smaller compared to $G_2^{Cu}$ and $G_B^{Pt}$ and thus can be ignored in Fig.~\ref{fig:spin_pumping_FMI_Cu_circuit}(c). Since $G_2^{Cu} \propto csch(t_{Cu}/\lambda_{Cu}) \simeq \lambda_{Cu}/t_{Cu}$, for $\lambda_{Cu} > 2t_{Cu}$, we get an inverse trend of $g_{eff,r}^{\uparrow \downarrow}$ with $t_{Cu}$ in the Fig.~\ref{fig:spin_pumping_FMI_Cu_results}.

To summarize, we have explained the critical trend of effective complex spin mixing conductance in magnetic insulator/heavy metal bilayers and explaining the experimental results with a sandwiched copper layer shows the prowess of the spin-circuit formalism. The interfacial spin mixing conductances for different samples of different thicknesses causes the variability in the experimental data points.~\cite{roy17_3} Similar analysis can be applied to other magnetic insulators and heavy metals in general.

\vspace*{2mm}
See the supplementary material for additional plots on thickness-dependent conductivity and spin diffusion length.

\vspace*{2mm}
This work was supported by Science and Engineering Research Board (SERB) of India via sanction order SRG/2019/002166.

\vspace*{1mm}
The data that support the findings of this study are available from the corresponding author upon reasonable request.

%\bibliographystyle{aipnum4-1}
%%%\nocite{*}
%\bibliography{royk,royk2}% Produces the bibliography via BibTeX.

\end{document}

\title{Supplementary Information\\Determining complex spin mixing conductance and spin diffusion length from spin pumping experiments in magnetic insulator/heavy metal bilayers}% Force line breaks with \\

\author{Kuntal Roy}
\email{kuntal@iiserb.ac.in}
\noaffiliation
\affiliation{Department of Electrical Engineering and Computer Science,\\Indian Institute of Science Education and Research Bhopal, Bhopal, Madhya Pradesh 462066, India}


% and this also matches the mathematical expression available in the literature.

\maketitle

\onecolumngrid

\begin{figure*}[h]
\centering
\includegraphics[width=0.4\textwidth]{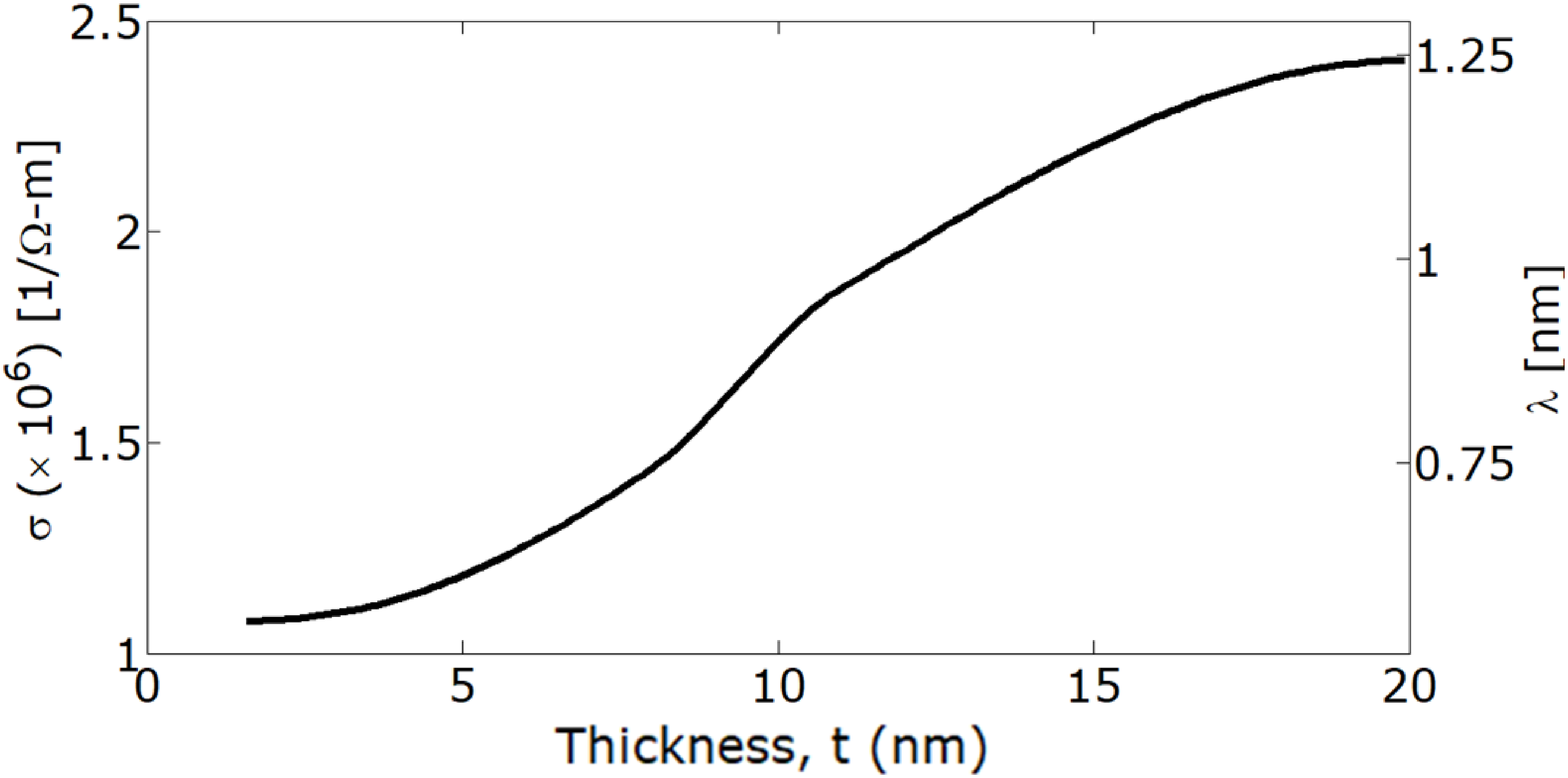}
\caption{\label{fig:spin_pumping_FMI_lambda} Thickness $t$ dependence of conductivity $\sigma$ and spin diffusion length $\lambda$ for the $YIG|Pt$ bilayer (Fig.~1 in the main Letter). Due to Elliot-Yafet spin relaxation mechanism, $\lambda(t) \propto \sigma(t)$. The dependence of $\sigma$ with thickness $t$ taken from Ref.~15 of the main Letter. The maximum value $\lambda_{max}$ ($\sigma_{max}$) = 1.24 nm (2.4e6 1/$\Omega$-m).}
\end{figure*}
%and the lowest value of $\lambda$ ($\sigma$) = 0.62 nm (0.73e6 1/$\Omega$-m) at $t=2$ nm

\begin{figure*}[h]
\centering
\includegraphics[width=0.4\textwidth]{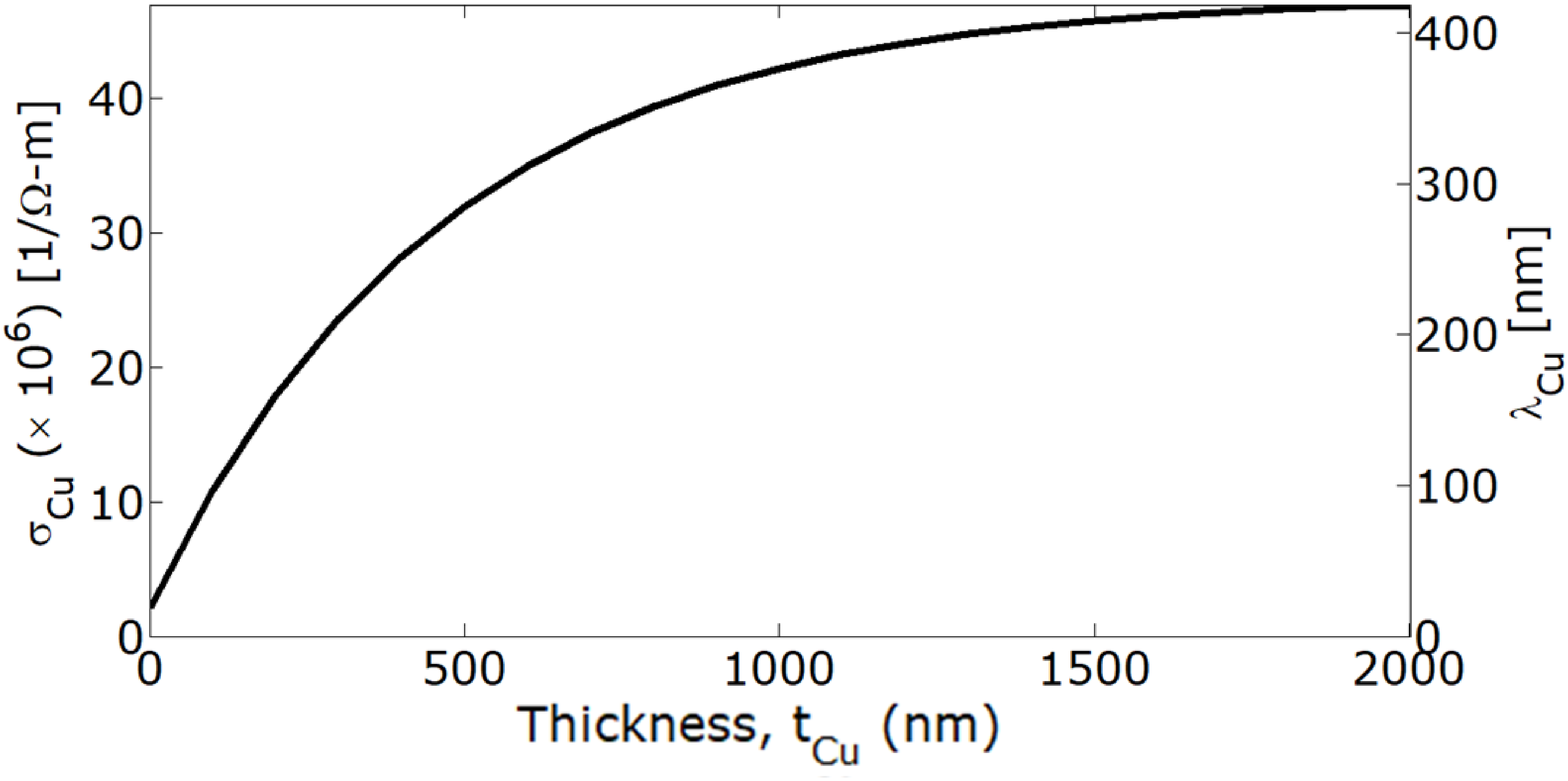}
\caption{\label{fig:spin_pumping_FMI_Cu_lambda} Thickness $t_{Cu}$ dependence of conductivity $\sigma_{Cu}$ and spin diffusion length $\lambda_{Cu}$ for the $YIG|Cu|Pt$ trilayer (Fig.~4 in the main Letter). Due to Elliot-Yafet spin relaxation mechanism, $\lambda_{Cu}(t_{Cu}) \propto \sigma_{Cu}(t_{Cu})$. The maximum value $\lambda_{Cu,max}$ ($\sigma_{Cu,max}$) = 418.4 nm (47e6 1/$\Omega$-m). Note that the maximum values of $\lambda_{Cu}$ and $\sigma_{Cu}$ used here are 32.58 nm  and 3.66e6 1/$\Omega$-m, respectively for the maximum fabricated thickness $t_{Cu}=17\,nm$.}
\end{figure*}